\documentclass{sigchi}

\toappear{
Permission to make digital or hard copies of all or part of this
work for personal or classroom use is granted without fee provided that 
copies are not made or distributed for profit or commercial advantage and
that copies bear this notice and the full citation on the first page. To
copy otherwise, or republish, to post on servers or to redistribute to 
lists, requires prior specific permission and/or a fee.\\
}

\usepackage{balance}  
\usepackage{graphics} 
\usepackage{times}    
\usepackage{url}      

\makeatletter
\def\url@leostyle{%
  \@ifundefined{selectfont}{\def\UrlFont{\sf}}{\def\UrlFont{\small\bf\ttfamily}}}
\makeatother
\def\pprw{8.5in}
\def\pprh{11in}

\usepackage{hyperref}
\hypersetup{
pdftitle={Crowds, Bluetooth and Rock'n'Roll: Understanding Music Festival Participant Behavior},
pdfauthor={},
pdfkeywords={Mobility, Bluetooth, Large-scale event, Festival},
bookmarksnumbered,
pdfstartview={FitH},
colorlinks,
citecolor=black,
filecolor=black,
linkcolor=black,
urlcolor=black,
breaklinks=true,
}

\begin{document}

\title{Crowds, Bluetooth and Rock'n'Roll:\\Understanding Music Festival Participant Behavior}

\numberofauthors{5}
\author{
  \alignauthor Jakob Eg Larsen \\
	\email{jaeg@dtu.dk}\\
  \alignauthor Piotr Sapiezynski \\
	\email{pisa@dtu.dk}\\
  \alignauthor Arkadiusz Stopczynski \\
	\email{arks@dtu.dk}\\
  \alignauthor Morten M{\o}rup \\
	\email{mm@imm.dtu.dk}\\
  \alignauthor Rasmus Theodorsen\\
	\email{ras.the@gmail.com}\\
  \end{tabular}\\
  \begin{tabular}{c}
	\affaddr{Technical University of Denmark}
}

\teaser{
  \centering
  \includegraphics[width=2.0\columnwidth]{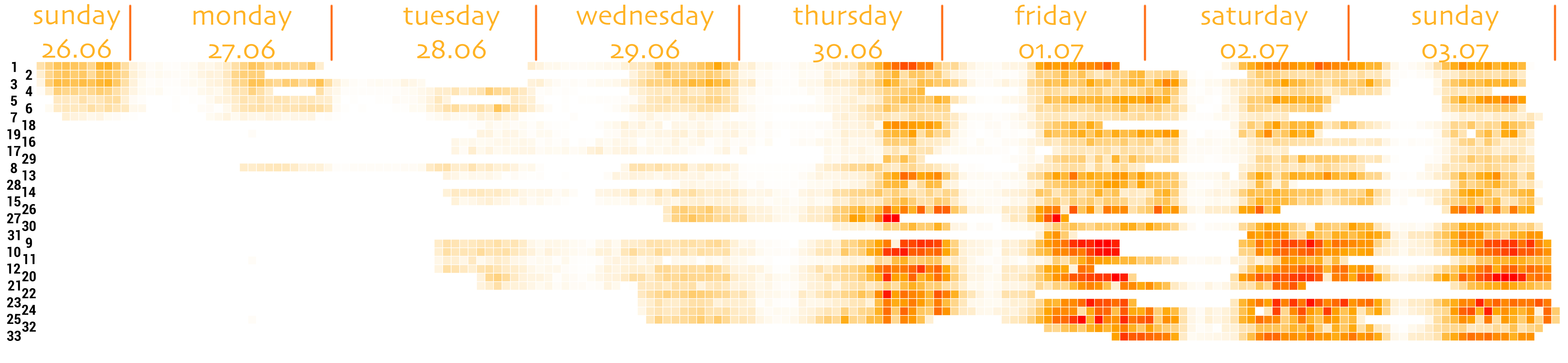}
  \caption{Unique Bluetooth devices observed throughout the 8 day festival by 33 proximity-based scanners, with the color intensity corresponding to the number of observations in one hour time windows. The scanners are grouped by stages and scanners at the main stages were deployed on day 4.}
  \label{fig:roskilde_day}
}

\maketitle




\begin{abstract}
In this paper we present a study of sensing and analyzing an offline social network of participants at a large-scale music festival (8 days and 130,000+ participants). 
Spatio-temporal traces of participant mobility and interactions were collected from 33 Bluetooth scanners placed in strategic locations at the festival area to discover Bluetooth-enabled mobile phones carried by the participants. 
We analyze the data on two levels. 
On the micro level, we use a community detection algorithm to reveal a variety of groups formed by the participants. 
On the macro level, we employ an Infinite Relational Model (IRM) in order to recover the structure of the network related to participants' music preferences. 
The obtained structure in the form of clusters of concerts and participants is then interpreted using meta-information about music genres, band origins,  stages, and dates of the performances. 
We show that the concerts clusters can be described by one or more of the meta-features, effectively revealing preferences of participants. 
Finally, we discuss the possibility of employing the described method and techniques for creating user-oriented applications and extending the sensing capabilities during large-scale events by introducing user involvement.
\end{abstract}




\section{Introduction}
Mobile phones have become increasingly ubiquitous and an integrated part
of our everyday life over recent years. This has led to a number of new
possibilities in studies of human mobility, behavior, and interactions,
as mobile phones can now be used to track people's activity. This area
has recently received increased attention with studies of mobility by
means of large phone data
sets~\cite{song2010limits,gonzalez2008understanding} or sensor data
collected on modern
smartphones~\cite{eagle2006reality,jensen2010estimating}. These studies
have reported insights into fundamental human mobility patterns with
results indicating very high levels of predictability.

In this paper we present a study of more than one hundred thousand music
festival participants mobility, group formation, and music preferences
at a large music festival in Denmark by using Bluetooth probing to
discover mobile phones carried by festival participants around the
festival area.

The use of Bluetooth technology as a way to gain insights into human
behavior and mobility has also received increased attention
recently~\cite{versichele2010potential}. Bluetooth technology has been
applied in several different domains, and different schemes have been
used.  In a study of mobility by Hui et al.~\cite{hui2005pocket}, 
participants were provided with a small
active Bluetooth device that they were carrying throughout a conference
to map participant mobility and events attended.
Most commonly Bluetooth scanners have been situated in fixed locations
to probe the presence of discoverable Bluetooth devices in proximity,
which is also the approach presented in this paper. This method has been
used for different applications including estimating the queue length
expressed in waiting time in airport security areas
~\cite{bullock2010automated,hansen2009location}. Large scale studies of
mobility by means of Bluetooth probing have also included tracking of
vehicles for the purpose of studying traffic
patterns~\cite{haseman2010real}, and large scale race events is another
example~\cite{stange2011analytical}.
In a related study O'Neill, Kostakos et al.~\cite{ONeill2006, Kostakos2010} concentrate on the mobility and interactions of participants with regard to semantic meaning of locations where the Bluetooth scanners were deployed. They show profoundly different patterns of presence in places of different social function, for example a busy street vs. a bar. Unfortunately, even though they deployed more than 90 scanners, they only refer to four categories of locations - a street, the university entrance, an office and a bar. It is not clear how much insight they gained into the social structure of other locations.

In the context of human mobility in festival settings, a study by
Versichele et al.~\cite{versichele2012use} also applies proximity-based
Bluetooth tracking to study mobility patterns. In their study 22
scanners were used over a duration of 10 days with 1.5 million
participants. However the general trend in their study is that
participants only visit the festival short-term (typically one day),
whereas the participants in this study are present at the festival area
for up to 8 days, and select among 160 music concerts and multiple other
events for the duration of the festival.

Where existing studies applying Bluetooth probing have focused on
describing mobility patterns, this study involves a richer semantic
context with information about concerts, music, genres, scenes, events,
and participants, allowing a more detailed contextual analysis of
participant behavior, and mobility.

More recently mobile sensor frameworks have been made
available~\cite{aharony2011social, kiukkonen2010towards} enabling the
collection of richer data sets capturing human behavior, mobility, and
data for mapping social interaction through multiple channels. An
advantage of having a mobile sensor framework on the smartphone is the
potential in combining multiple sensor data to obtain finer granularity information and 
more robust estimations.
For instance, data from sensors such as GPS,
WiFi, GSM, and accelerometer can be combined
to build a location estimator which works in different contexts (outdoors and inside
the buildings) with higher accuracy than any single of these sensors can provide
~\cite{montoliu2010discovering}. However, a
challenge in these studies is the deployment which involves a mobile
application running on participant devices. Therefore these studies have
typically been carried out on a smaller selected population, but often
over longer periods of time. As a result, the observational conditions
and especially population sampling may introduce unknown biases.
Although a mobile client (smartphone) may lead to very rich data sets,
this methodology has a different set of challenges in terms of
deployment at the festival. This includes supporting multiple clients
and that participants have to actively install an application containing
mobile sensing components.  In this study the duration of the event is
only 8 days, but using the Bluetooth probing technique we have access to
a larger population.

In the following sections we descrbie the methodology, limitations, and
challanges of data collection using Bbluetooth scanning system in an
environment with a limited and short-lived technical infrastructure.
Next we present the data acquired during the 8 days of the festival and
discuss the results of the Bluetooth discovery process. The chapter is
concluded with a discussion of potential applications and the insights
that can be obtained from studying the spatio-temporal data that can be
acquired through Bluetooth probing. 
\section{Roskilde Festival}
Roskilde Festival is one of the six biggest annual music festivals in
Europe and is held south of Roskilde in Denmark. It started in 1971 and
since 2009 has been attracting more than 100,000 participants annually
(with up to 30\% being volunteers). In 2011 it gathered an estimated
130,000+ people. The festival lasts for 8 full days, starting Saturday
evening and finishing Sunday at midnight. For the first 4 days only the
camping grounds and a small festival zone are open, including a single
stage (Pavilion Junior) featuring upcoming Nordic bands. On Thursday
afternoon the main grounds are opened, the major music
events start and last for the next 4 days.

The main festival grounds cover about 0.2 km$^{2}$ with 6 stages of
various sizes. The festival campsite, located south to the main festival
grounds covers nearly 1 km$^{2}$. In addition to the stages, the grounds
include cultural zones, shops, restaurants, artistic installations etc.
Participants can freely move through the grounds in the day time; once
the concerts are finished for the day, the main grounds are closed and
then open in the morning hours next day. Some areas in the main grounds
are off limits for participants, such as backstage areas or technical
areas behind merchandise passages.

In 2011 the participants consisted of 77,500 festival guests, around
3,000 press representatives, 3,000 artists, 30,000 volunteers, 20,000
one-day guests plus an unknown number of guests over 60 and under 10
years old -- we estimate that at least 130,000 people present at the
festival during the 8 days in total. 54\% of the population were women and approx
22\% of the audience visited the festival for the first time. The
average age was 23 years and a typical participant was a student living
in a Scandinavian city. 80\% of the participants came from Denmark, 8\%
from Norway, 4\% from Sweden, and 8\% from other
countries\footnote{Source: \url{http://roskilde-festival.dk/}}.

The six stages host concerts of different sizes and genres:
\begin{itemize} \item Orange stage, capacity 60,000+, all genres \item
Arena stage, capacity 17,000, all genres \item Cosmopol stage, capacity
6,000, hip-hop, electronica, urban world music \item Odeon stage,
capacity 5,000, mixed, mostly rock \item Pavilion stage, capacity 2,000,
mixed, mostly rock \item Gloria stage, capacity 1,000, mixed,
experimental \end{itemize}
\section{Methodology}
Our study of human mobility in the festival settings relies on discovering
Bluetooth-enabled devices that are operating in discoverable mode. As
Bluetooth is a short-range low-power protocol for implementing Wireless
Personal Area Networks (WPAN), it limits the range in which
Bluetooth-enabled devices can be discovered. It operates on the
Industrial, Scientific and Medical (ISM) frequency band of
2.4GHz~\cite{peterson2006bluetooth}. Communication always happens in
master-slave mode and is established between new devices with a master
device sending inquiry packets to discover nearby devices that are in
the inquiry scan substate (discoverable). Discoverability of a device
commonly needs to be set manually by the user, and can be either limited
in time or set to infinite. It is worth noticing that for instance
Android-based smartphones (until recent versions) only allow time
limited discoverability, while iOS devices (iPhone, iPod, etc.) and WindowsPhone
smartphones are only discoverable while the user is interacting with
the Bluetooth menu. While this limits the number of potential phones
we can discover significantly, we show that there are still many discoverable devices.

In the present study Bluetooth scanners functioned as master devices,
broadcasting inquiry messages (scanning) continuously. Responses from
the devices in proximity were silently logged, without any active
participation on the user side. This is similar to the approach
described in~\cite{hay2009bluetooth} where tracking of the individual in
a non-invasive way is considered more suitable for large-scale studies.
The received signal strength intensity (RSSI) of the response was not
registered. Although it is technically possible to use RSSI to calculate
the position of the discovered device through
multilateration~\cite{bensky2007wireless, kelly2010minimal}, the
accuracy of the approach varies depending on the environment. Moreover,
due to the limited range of Bluetooth, we considered position accuracy
obtained from a single scanner (i.e.~around 10 meter radius for class 2
Bluetooth devices) sufficient.

\subsection{Bluetooth Scanner Device}

Off-the-shelf Nokia N900 smartphones were used as Bluetooth scanners
with custom software built for detecting Bluetooth-enabled devices in
proximity. Off-the-shelf hardware was used as a relatively simple
solution, providing 3G communication (necessary for obtaining the
results in real time from the large festival area), data storage,
battery power (for the events of short power outages), GPS for tracking
the device in case it was lost, and finally a Bluetooth module. The data
from the scans was stored in a local SQLite database on the device and
additionally uploaded to a server, depending on the network
availability. Scanner and uploader applications were running on the
smartphone, and extra background processes restarted them if required.
This was to ensure the highest possible availability and robustness of the system.

A scan for discoverable devices typically takes about 30 seconds, so
scanning performed as frequent as possible results in approximately two
scans per minute. Devices that did not upload data to the server for a
prolonged period of time were rebooted either by issuing a command via
Bluetooth or by manually turning them on and off. In order to minimize
this effect, periodical reboot every 24 hours was enforced in the
software.

The collected data is a time-series of events. Each of the events is
described by the time, scanner ID and a Bluetooth MAC address of 
a discovered device. This information does not enable us to link the device
to the person (such as name or personal identification). Thus, the Danish
Data Inspectorate considered the information handled in this project as
being non-sensitive information about the participants thereby enabling
the observations to be made without special permissions or requiring informed
consent from the participants. To ensure that
not even the detected devices were identifiable afterwards, the MAC
addresses were hashed after extracting information about the vendor. The
human-readable identifiers (Bluetooth friendly names) of the devices
were not retrieved in order to improve the scanning time and to ensure
anonymity of the participants. 
\section{Observational Study}
The data was captured through 33 Bluetooth scanners placed in strategic
positions around the festival site, as shown in
Figure~\ref{fig:scanners}. The scanners were placed in the vicinity of
the stages, as those were the most interesting, semantically rich spots.
However, since the availability of power sources was crucial while choosing the 
exact location, and the infrastructure at the festival is only temporary, 
the scanners were mainly located in the shops, beer booths 
(close to the counters) and mixing areas of the stages. Those locations 
provided sufficient coverage of relevant areas to discover patterns in
participants' mobility. 

\begin{figure}[!h] \centering
\includegraphics[width=1.0\columnwidth]{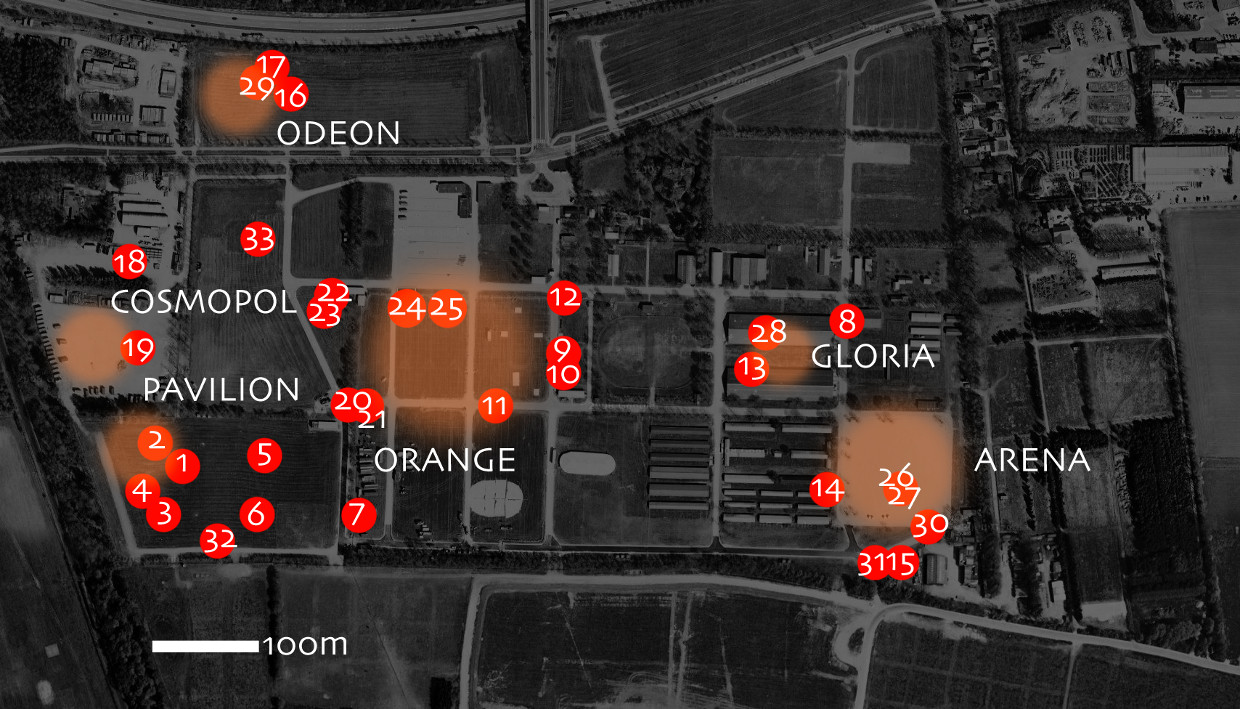}
\caption{Map of Roskilde Festival inner area with indication of the
location of Bluetooth Scanners. The orange areas indicate places for the
audience for the respective stages.} \label{fig:scanners} \end{figure}

The Bluetooth scanning data was uploaded in real time via the 3G network
for real time processing purposes, but upload of the data was of course
subject to network availability. Problems with the mobile network
connections were occurring due to a high number of mobile phones in a
relatively small physical area, especially during large concerts.
Therefore, for some scanners the collected data was uploaded once the
connection was available (typically in early morning hours). 7 of the 33
devices were running without manual intervention for the whole period of
the festival, but the rest had to be maintained one or more times during
the festival. For instance, if the power had been switched off for more
than about 7 hours (typically in the early morning) the devices had to
be manually turned on.

The radius of Bluetooth is limited to about 10 meters for the
transmitters used in most of the mobile phones (class 2). This makes it
possible to pinpoint the location of the observed devices, however
making it a challenge to collect representative data in a large area, as
it will only be partly covered. The devices observed by a scanner could
belong to a person only passing by; on the other hand, a person staying
right outside the radius of the coverage even for the whole concert
might not be discovered.

\begin{figure}[!h] \centering
\includegraphics[width=0.5\columnwidth]{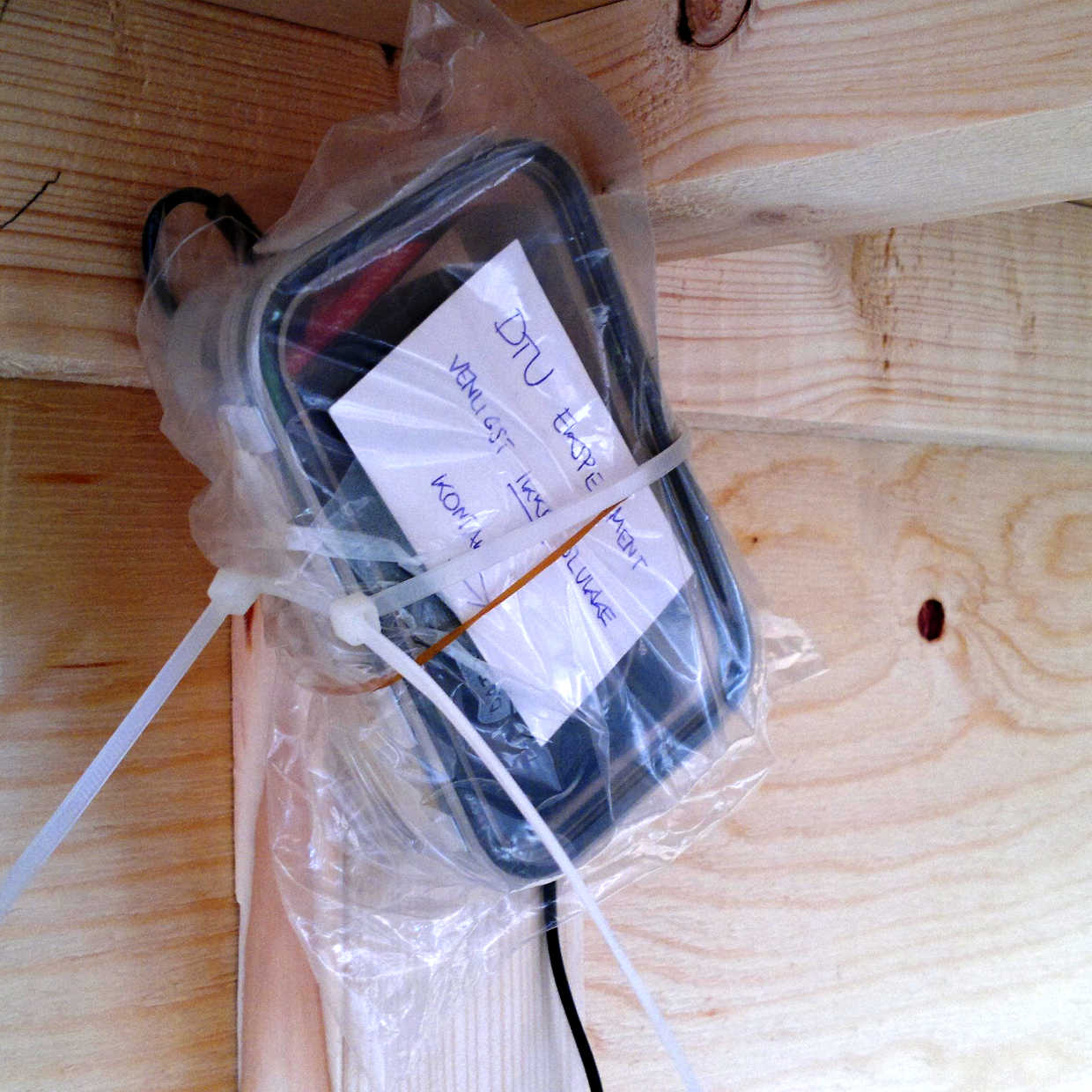} \caption{Nokia
N900 Bluetooth scanner in a protective box attached under a beer booth
counter.} \label{fig:scanner} \end{figure}

\section{Data Collection and Analysis}

The deployed Bluetooth scanners collected 1,204,725 observations during the 8 days of festival activities. This included a total of 8,534 unique devices discovered, meaning an average of 141 observations of each device during the festival. Overall, this corresponds to at least 6.5\% of the population at the festival have been observed in the study, thereby providing a window to understand festival participant behavior, mobility, and interactions. 

Table~\ref{tab:table_scanners} provides an overview of the observations from the 33 scanners used in the study.  As can be seen from the table the most unique devices were discovered by scanner 9, 10, 23, 24, and 25 that are all located around the largest stage where most participants would be expected to be seen. Each of those scanners discovered above 4,000 unique devices. An overview of unique devices observed throughout the 8 days of the festival is shown in Figure~\ref{fig:roskilde_day} on the first page.

\begin{table}
  \centering
\begin{tabular}{|c|c|c|c||c|c|c|c|}
\hline
\#&Obs&Uniq&O&\#&Obs&Uniq&O\\
\hline
1&77145&3607&1&18&28844&2302&3\\
2&44224&1880&9&19&32773&2245&8\\
3&53706&3091&11&20&34264&4753&15\\
4&31836&1801&15&21&22022&3473&20\\
5&33167&3265&16&22&20003&1901&2\\
6&38834&2120&20&23&43784&4372&29\\
7&28440&1102&3&24&53695&4404&27\\
8&40648&893&0&25&55025&4429&51\\
9&49852&4316&2&26&61706&3290&12\\
10&45813&4116&28&27&24714&1900&16\\
11&21714&3467&3&28&32512&1651&8\\
12&30027&3433&31&29&27944&1491&5\\
13&60276&2770&11&30&32067&2411&22\\
14&34202&3159&8&31&15616&2514&21\\
15&36293&2582&5&32&19190&2553&22\\
16&22044&1809&3&33&25578&2934&18\\
17&20280&1227&13& & & & \\
\hline
  \end{tabular}
  \caption{An overview of the 33 scanners with numbers of observations (Obs), and unique devices per scanner (Uniq), and unique devices only discovered per scanner (O). Total number of unique devices was 8,534.}
  \label{tab:table_scanners}
\end{table}

Beyond serving as a unique identification of the device the MAC address is structured so the vendor of the device can be determined from first three octets of the address (24 bits) formally known as an "Organizationally Unique Identifier" (OUI)~\cite{ieeeoui}. The list of the assigned OUIs is managed by IEEE, designated by the ISO Council to act as the registration authority. Some identifiers found in the devices may not correspond directly to the end-product manufacturers, as they may be registered under subcontractors company. In total, around 70 unique vendors were discovered, however the 7 largest vendors account for 96\% of all unique devices and 99\% of all observations, as shown in Figure~\ref{fig:unique_devices}. 

\begin{figure}[!h]
\centering
\includegraphics[width=0.9\columnwidth]{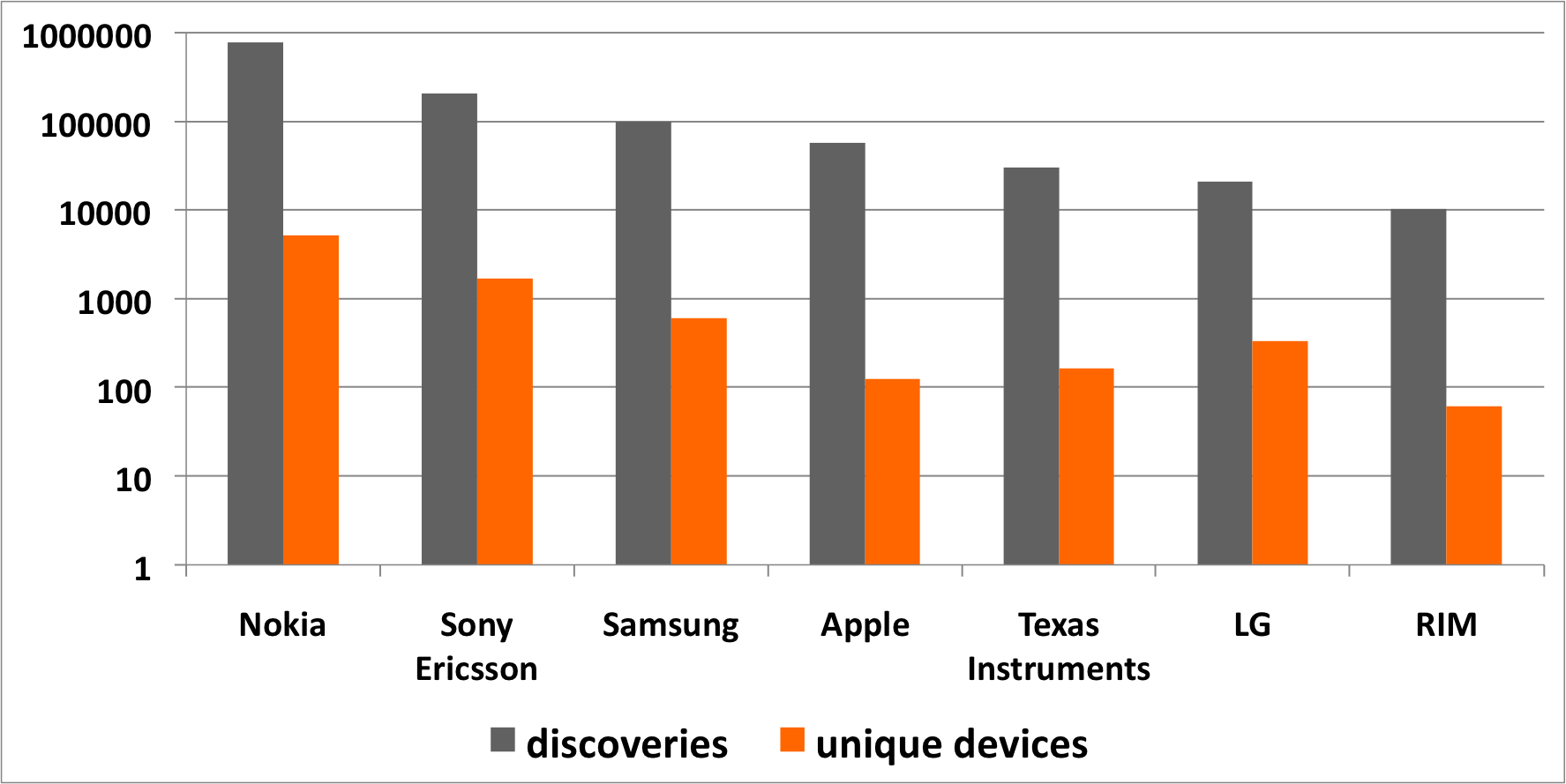}
\caption{Number of unique observations and devices (log scale). The 7 largest vendors account for 96\% of the devices and 99\% of the observations.}
\label{fig:unique_devices}
\end{figure}

\section{Modeling}
One of the interesting questions regarding events such as music festivals is the internal structure of the crowd: whether people move alone or in groups and how groups are different. In addition, the influence of music taste on collective group decisions on concert selections is interesting as is the mobility of the groups. In attempt to understand what insights on these issues we can gain from the data obtained in the presented study, we analyze the data at two levels. Firstly, we concentrate on the micro level by running a community discovery algorithm. Then, we investigate the macro level to determine the general trends of attendance and relate the findings to the available meta information regarding the schedule of the festival and types of artists.
\subsection{Micro Groups Modeling}

We understand micro groups as sets of people frequently co-occurring in spatio-temporal bins. We divide the timeline of the entire festival into 1076 x 10-minute temporal bins, 10 meter radius areas around the scanners create the spatial bins, as shown in Figure~\ref{fig:scanners}. A similar technique of inferring social links from spatio-temporal co-occurrences is described in~\cite{Crandall28122010}.

Out of 8,534 discovered devices we rejected those, which were seen in less than 10 temporal bins or less than 3 spatial-bins. Those devices were considered belonging to participants for whom we do not have sufficient data or being stationary devices (such as crew laptops). After this processing 5,339 devices were obtained (63\%). For all these common co-occurrences were calculated.  The weights of the links were calculated as the number of co-occurrences of participant A with participant B divided by total number of occurrences of participant A (A to B edge). This creates a directed graph, where A can be important to B but not necessarily the other way around. This accounts for the asymmetry in the participants’ activity and different natures of their relations. For the visualization and subsequent analysis, only links that occurred in at least 3 different locations and weight at least 0.5 (seen in 50\% of all observations of the participant) were chosen. 

\begin{figure}[!h]
\centering
\includegraphics[width=1.0\columnwidth]{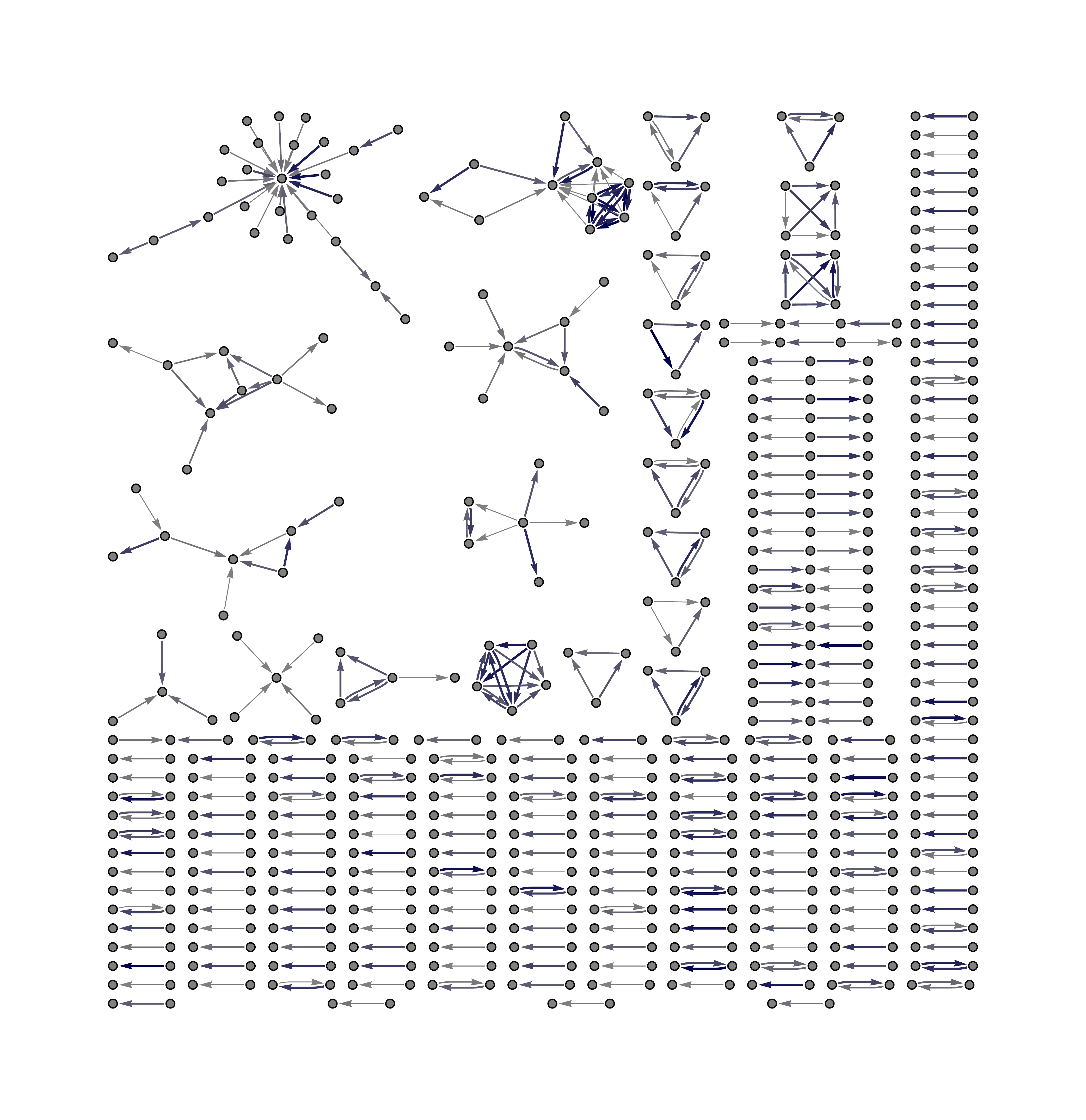}
\caption{Directed graph of discovered micro-groups of participants. Participants frequently seen together create couples, triangles and larger structures, providing insights into the internal structures of the festival crowd.}
\label{fig:micro_groups}
\end{figure}

The final constraints on the discovered micro groups are strong: they require that from 130,000 participants, people are seen within 10 minutes in a radius of approx.~10 meters at least half of the times they are observed in total and in at least 3 different locations, to ensure sufficient entropy for meaningful modeling. The constraint are imposed on existence of each edge, hence the directed edges. This should ensure that the discovered motifs are in fact people moving around together. It was found that 12 nodes (devices) were forming structures (pairs and square) with perfect correlation of occurrences which we consider devices belonging to the same person. Based on the discovered groups, a directed graph can be constructed, with edges indicating the discovered friendships. In total, 574 nodes with 448 edges were detected. The motifs can be seen in Figure~\ref{fig:micro_groups}. The most interesting are the structures with high connectivity, indicating groups of participants observed to often move together.

The baseline for the micro group detection was calculated using rewiring algorithm \cite{maslov2004detection}, shuffling the participants in spatio-temporal bins. For N=35 tests $\mu = 5$ nodes and $\mu = 3$ edges were discovered ($\sigma = 4.32$ and $\sigma = 2.60$ respectively). This indicates that the recovered structures are not an effect of random movement of participants but reflect an actual underlying structure.

The star structure visible in the upper left corner of Figure~\ref{fig:micro_groups} with multiple inbound edges and none outband is an interesting artifact showing a person working in a shop in an area covered with several scanners. The person was frequently picked up by 3 scanners (1, 2, and 3) with customers also picked up there but independently from each other. Similar artifacts were seen in larger number when the threshold of common co-occurrences was set to 2, since some of the long beer booths had two scanners placed in them. Such star structures with inbound links can be summarized by saying that those places (represented by people working in them) were important to participants, but participants were not significant for them.

The presented algorithm for detecting micro groups and discovered structures is a simple example of possibly very granular analysis of the collected data. With extremely small spatio-temporal bins we still recover over 500 people moving around while belonging to a particular structure.
\section{Macro Groups Modeling}
We combine the spatio-temporal traces with the band schedule, to find out which concerts each of participants attended. Next, we assign a set of meta information to each show. This way we establish a richer semantic context and analyze the guests' motivations for choosing particular concerts. The metadata consists of:

\begin{itemize}
\item {\em genre} -- based on available Last.fm tags, each band is manually assigned with one genre label from the following: {\em electronic}, {\em rock/pop}, {\em folk/world}, {\em hip-hop/rap}, {\em metal/punk/hardcore}, {\em other}
\item {\em playcount} -- number of times Last.fm users listened to music of a band
\item {\em country of origin} -- from the Roskilde Festival schedule; the countries have been grouped into following categories: {\em Denmark}, {\em Other Nordic}, {\em USA}, {\em Western Europe}, {\em Other}
\item {\em scene} -- from the Roskilde Festival schedule
\item {\em date} -- from the Roskilde Festival schedule
\end{itemize}

Intuitively, the number of people at the concert would be highly correlated with the intensity with which people listen to the bands, i.e. the playcount. To verify this assumption, we calculate Pearson's correlation between the number of unique devices found during each concert and the logarithm of playcount of the band, see Table~\ref{tab:tab_correlation}. We group the concerts according to the size of the stage they performed at. As shown in Table~\ref{tab:tab_correlation} there is a small (if any) positive correlation between the popularity of the band and the number of discovered devices. This shows that people's choices regarding the concerts they attend can not be fully accounted for in this way and more complex modeling should be used to reveal more interesting patterns.

\begin{table}
\centering
\begin{tabular}{| c | c | c | c |}
\hline
Size of stage & Small & Medium & Big \\ \hline
$\rho$ & 0.2462 & 0.0351 & 0.3427 \\ \hline
$P-Value$ & 0.0333 & 0.8593 & 0.0091 \\ \hline
\end{tabular}
\caption{Correlation between popularity of the band (log playcount) and the number of unique devices}
\label{tab:tab_correlation}
\end{table}

	\subsection{Data pre-processing}
Our Bluetooth traces are a time-series of events, each of which contains the participant id, scanner id, and time. The goal of the pre-processing stage is to transform the behavioral time-series data into a binary attendance table, which maps each participant to the concerts she attended. In each event, we assign the scanner to the stage where it was located. Then, we assume that scans which took place between 10 minutes before the starting time of a concert and 1 hour 45 minutes after that moment were taken ``during'' this concert. Thus, we determine during which concert, if any, each event happened. This results in a matrix where each element represents the number of times each participant was scanned at a given concert. To indicate whether a given participant actually attended a concert, we transform the table to a binary table by setting a threshold on the number of observations.

	\subsection{Outlier detection}
The binary table created in pre-processing contains two categories of outliers. Firstly, there are guests who participated in less than three concerts and are thus irrelevant in terms of the analysis. Bluetooth devices, which were recorded throughout the festival at the same location such as employee cell phones or laptops at a particular stage constitute the second category of outliers. These are defined as entities which participated in at least 70\% of concerts at one stage and at least in twice as many concerts at that one stage compared to all the other stages in total. After removing outliers, 5127 attendees are left for further analysis.

	\subsection{Metadata pre-processing}
We obtain the community assigned tags for each band from Last.fm. There are more than 400 unique tags associated with the participating bands and for our modeling purposes we need to significantly reduce the dimensionality of this data. Based on the most significant tags and manual verification, we assign each band to one particular genre: {\em electronic}, {\em rock/pop}, {\em folk/world}, {\em hip-hop/rap}, {\em punk/metal/hardcore}, {\em other}. Such categorization is, of course, highly simplified, but provides a satisfactory representation of kinds of music performed at the Roskilde Festival.


	\subsection{The Infinite Relational Model}
We fit an Infinite Relational Model\cite{Kemp2006,Xu2006} to the binary attendance matrix to reveal the underlying patterns of people's behavior at the festival. Note, that the Model is oblivious to the accompanying meta information such as genre, band's country of origin, date, and location of each show.

\newcommand{\CRP}{\mathrm{CRP}}
\newcommand{\Beta}{\mathrm{Beta}}
\newcommand{\BetaInc}{\mathrm{BetaInc}}
\newcommand{\B}{\mathrm{B}}
\newcommand{\G}{\mathrm{G}}
\newcommand{\Bernoulli}{\mathrm{Bernoulli}}
\newcommand{\Poisson}{\mathrm{Poisson}}
\newcommand{\Gam}{\mathrm{Gamma}}
\newcommand{\GamInc}{\mathrm{GammaInc}}
\newcommand{\m}[1]{\boldsymbol{#1}}

The \emph{infinite relational model} (\textsc{irm}) is a model for binary relational data (graphs) and can be characterized by the following generative process for bipartite graphs. First, each of the row and column nodes are assigned to a cluster according to the Chinese restaurant process ($CRP$). The $CRP$ is an analogy for building a partition ground up by assigning the first node (i.e.~customer in a restaurant) to a table and subsequent nodes (customers arriving at the restaurant) to an existing table, i.e.~cluster, with probability proportional to how many existing customers are placed at the table and at a new table, i.e.~cluster, with a probability proportional to the parameter $\alpha$. Customers thereby tend to sit at most popular tables making the popular tables even more popular -- an effect noted as the rich gets richer. The partition of the nodes induced by the $CRP$ is exchangeable in that the order in which the customers arrive does not influence the probability of the partition\cite{pitman2006combinatorial}.
Next, link probabilities are generated which specify the probability of observing a link between clusters; and finally, the links in the network are generated according to these probabilities. For bipartite graph we have the following generative process:

\begin{align}
\m z^{(1)} & \sim \CRP(\alpha^{(1)}), & & \textit{Row cluster assignment,}\\
\m z^{(2)} & \sim \CRP(\alpha^{(2)}), & & \textit{Col. cluster assignment,}\\
\eta_{\ell m} & \sim \Beta(\beta,\beta), & & \textit{B/t. cluster link prob.,} \\
A_{ij} & \sim \Bernoulli(\eta_{z_i^{(1)}z_j^{(2)}}), & & \textit{Link.}
\end{align}

Inference in the \textsc{irm} model, i.e.~determining the posterior distribution of the cluster assignments, entails marginalizing over the link probabilities, which can be done analytically. This is a major advantage of the \textsc{irm} model, enabling inference by Markov chain Monte Carlo (MCMC) sampling over the cluster assignments alone.
Marginalizing over link probabilities, i.e.~$\m \eta$, we obtain the following joint posterior likelihood

\scriptsize
\begin{align*}
p(\m{A},\m{z}^{(1)},\m{z}^{(2)}|\beta,\alpha^{(1)},\alpha^{(2)})&=&
p(\m{A}|\m{z^{(1)}, z^{(2)}},\beta)
p(\m{z}^{(1)}|\alpha^{(1)})
p(\m{z}^{(2)}|\alpha^{(2)})\\
&=&
\Bigg[\prod_{\ell m} \frac{\Beta(N^{+}_{\ell m}+\beta,N^{-}_{\ell m}+\beta)}{\Beta(\beta,\beta)}\Bigg]
\times\\
&& \Bigg[\frac{\alpha^{(1)^L}\Gamma(\alpha^{(1)})}{\Gamma(I+\alpha^{(1)})}\prod_{\ell=1}^{L^{(1)}}\Gamma(M_\ell^{(1)})\Bigg]
\cdot\\
&& \Bigg[\frac{\alpha^{(2)^L}\Gamma(\alpha^{(2)})}{\Gamma(J+\alpha^{(2)})}\prod_{\ell=1}^{L^{(2)}}\Gamma(M_\ell^{(1)})\Bigg],
\end{align*}
\normalsize

where $L^{(k)}$ is the number of clusters, $M_\ell^{(k)}$ is the number of nodes in the $\ell$th cluster of mode $k$, and $N^+_{\ell m}$ and $N^-_{\ell m}$ are the number of links and non-links between nodes in cluster $\ell$ and $m$.
Using Bayes theorem the conditional distribution of the cluster assignment of a single node is given by

\scriptsize
\begin{align*}
p(z_i^{(1)}\!=\!\ell|\m A,\m z^{(1)}\!\!\setminus\!\!z_i^{(1)},\alpha^{(1)},\m z^{(2)}, \beta)
\propto \Bigg[\prod_{m} \frac{\Beta(N^{+}_{\ell m}+\beta,N^{-}_{\ell m}+\beta)}{\Beta(N^{+\setminus i}_{\ell m}\beta,N^{-\setminus i}_{\ell m}\beta)}\Bigg]
q^{(1)}\\
p(z_j^{(2)}\!=\!m|\m A,\m z^{(2)}\!\!\setminus\!\!z_j^{(2)},\alpha^{(2)},\m z^{(1)},\beta)
\propto \Bigg[\prod_{\ell} \frac{\Beta(N^{+}_{\ell m}+\beta,N^{-}_{\ell m}+\beta)}{\Beta(N^{+\setminus j}_{\ell m}\beta,N^{-\setminus j}_{\ell m}\beta)}\Bigg]
q^{(2)},
\end{align*}
\normalsize

such that $q^{(k)}=\left\{\begin{array}{cc}
w^{(k)}_\ell &if \ w^{(k)}_\ell>0\\
\alpha^{{(k)}} &otherwise\end{array} \right.$ where $w_{\ell}$ is the number of nodes already assigned to cluster $\ell$ and $N^{+\setminus i}_{\ell m}$ and $N^{-\setminus i}_{\ell m}$ denotes the number of links and non-links between nodes in cluster $\ell$ and cluster $m$ not counting any links from node $i$ of mode one ($j$ is similarly used to denote not counting any links from node $j$ in mode two).
Hence, a new cluster is generated according to the $\CRP$ with probability proportional to $\alpha^{(k)}$.
By (Gibbs) sampling each node assignment of the row ($z_i^{(1)}$) and column ($z_j^{(2)}$) clusters in turn from the above posterior distribution we can infer $\m z^{(1)}$ and $\m z^{(2)}$. The inference thereby also estimates from data the number of groups in each mode.

We note that this posterior likelihood can be efficiently calculated only considering the parts of the computation of $N^+_{\ell m}$ and $N^-_{\ell m}$ as well as evaluation of the Beta function that are affected by the considered assignment change. Notice, the expected value of the relations $\boldsymbol{\eta}$ given the node assignments $\boldsymbol{z^{(1)}}$ and $\boldsymbol{z^{(2)}}$ is defined by $\langle\boldsymbol{\eta}_{lm}\rangle=\frac{N_{lm}^++\beta}{N_{lm}^+N_{lm}^-+2\beta}$. Apart from the above Gibbs sampling we also include so-called split-merge moves to improve the inference \cite{Jain2004}. The split merge procedure was implemented with three restricted Gibbs sampling sweeps initialized by the sequential allocation procedure of \cite{Dahl2005}. \emph{Infinite relational model} can be efficiently applied to large datasets using GPU computing \cite{hansen2011}, which could allow for real time applications. Here we set $\beta=1$, $\alpha^{(1)}=log(I)$ and $\alpha^{(2)}=log(J)$, where I is the number of unique devices and J is the number of concerts.

	\subsection{Robustness of the model}

We use a number of measures to evaluate the generalizability of the results and robustness of the model. The model estimation procedure is run 110 times; each time 2.5\% of the links and an equal number of non-links are treated as missing, and then used for prediction. Firstly, normalized mutual information ($NMI$) is calculated between each pair of estimated models. Notice, $0 \leq NMI \leq 1$ where 0 indicates no relationship between the two assignment matrices and 1 indicates a perfect correspondence \cite{hansen2011}. The $NMI$ scores for the concert assignment matrices average at 0.91 with the standard deviation of 0.03, while the score for the attendee assignment matrices has the mean of 0.45 and standard deviation of 0.02. The relatively low $NMI$ for the clusters of participants is related to the fact that the model forces the assignment of each attendee to only one cluster. There can be many such assignments which are equally valid and thus with every run of the model calculation the final participant groups vary. Since the assignments of concert clusters are significantly more stable, they will be in focus of further analysis.

The predictive performance of the model is measured using the Area Under Curve ($AUC$) of the Receiver Operator Characteristic. $AUC$ evaluates how well the distributions of links and non-links are separated. Notice, $0\leq$AUC$\leq1$ where 0.5 indicates separation not better than a random guess and 1 indicates a perfect separation. This measure is not vulnerable to class imbalance problem \cite{steinbach2006}. The average value of $AUC$ for the 110 models is 0.81 with the standard deviation of 0.01. Finally, it is shown that after 150 iterations the log probability of the model converges to a stable value across 110 runs, see Figure~\ref{fig:robustness}. It is important to emphasize that this stability is achieved for the models trained on non-complete datasets (with each run 2.5\% of links and the equal number of non-links were randomly discarded to be used for prediction). As shown in Figure~\ref{fig:robustness} the model is robust to random initialization conditions as well as to data partially missing.

\begin{figure}[!h]
\centering
\includegraphics[width=0.9\columnwidth]{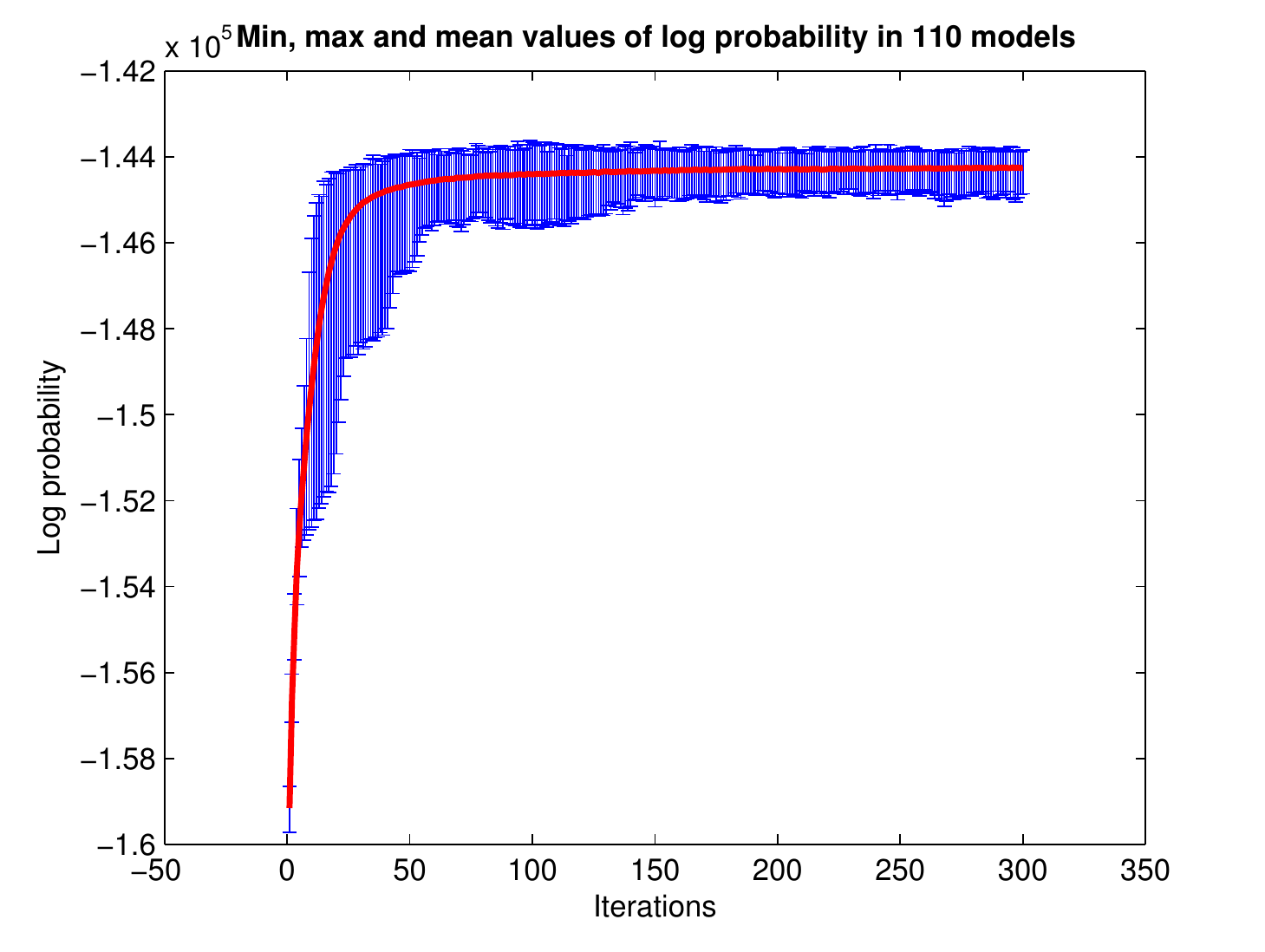}
\caption{Robustness: Independently of random initialization conditions and parts of the data used for cross-validation, the final value of log likelihood is stable across 110 trained models.}
\label{fig:robustness}
\end{figure}

	\subsection{Results}
After having proven the stability and generalizability of the used method, more models are calculated based on the full attendance table, without treating any part of the data as missing. The model with highest log probability is used for further investigation. As shown in Figure~\ref{fig:concert_user} this model groups 5127 people in 16 clusters and the 160 concerts in 25 clusters. The color coded value of $\eta$ indicates the between-cluster link probability. In subsequent sections these values are interpreted and related to the available meta information.

\begin{figure}[!h]
\centering
\includegraphics[width=0.9\columnwidth]{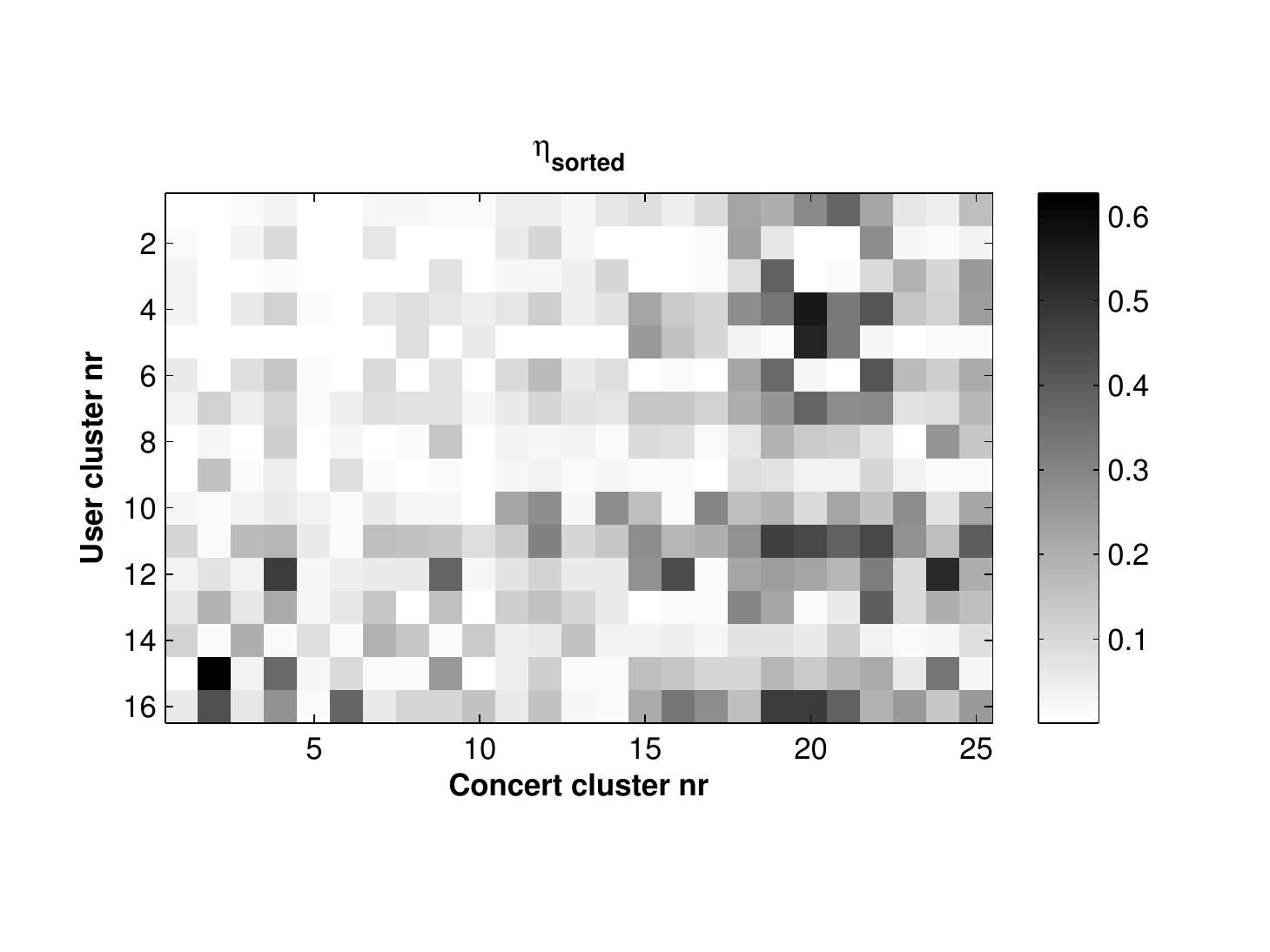}
\caption{Between cluster link probability for the estimated 16 clusters of attendees and 25 clusters of concerts, with clusters sorted by size in descending order. Preference regarding the choice of concerts can be observed, for example user cluster 5 is strongly associated with concert clusters 15, 16, 17, 20, 21 – many people in cluster 5 attended concerts from these clusters.}
\label{fig:concert_user}
\end{figure}

		\subsubsection{Relating chosen concert clusters to available metadata}

This section describes particular findings which further justify the use of the chosen technology as well as provide additional insight into the audience dynamics.

Figures \ref{fig:clusters_date} - \ref{fig:clusters_summary} show the distribution of concerts in the created clusters in relation to particular features. We only consider first 10 clusters, containing between 24 (cluster 1) and 7 (cluster 10) concerts. This captures .725 of all concerts at the festival. With less concerts in clusters, it is increasingly hard to provide meaningful interpretation.

We use $\chi^2$ test to compare the distributions in the clusters against the overall distribution to understand if the cluster bears any meaning in relation to the particular feature. It should be however emphasized that the results are not rock-solid: with such a small number of concerts in the clusters, the results are more of a guidance in relating the clusters to available metadata, rather than quantification of the findings. Still, we can note that the model produces interpretable results, giving insight into the festival structure.

Figures \ref{fig:clusters_date} - \ref{fig:clusters_stage} show the distribution of concerts from clusters (1-10) across the available meta information. The last column in each figure indicates whether the distribution in that cluster is significantly different than the overal distribution: if yes, the cluster can be considered meaningful and explained by this feature. Figure \ref{fig:clusters_date} shows that the clusters are quite structured in terms of the dates. It is intuitively understood - the concerts are attended by festival participants present at that particular day. As shown in Figure \ref{fig:clusters_genre} only two clusters have distribution of genres different than overall distribution. These two clusters clearly point to electronic and folk/world genres. Figure \ref{fig:clusters_origin} deals with the distribution of origin of the bands and shows three clusters with well-pronounced grouping of the bands: Danish, Danish+Nordic, and USA. 

Figure \ref{fig:clusters_stage} indicates that most the clusters display strong grouping of the bands based on the stage where they happened. This may be related to the fact that concerts of similar type (if not necessary the same genre) are planned at the same stage; also, participants mobility is limited and a common behavior of participants may be to stay at the same stage.

The summary shown in Figure \ref{fig:clusters_summary} makes it clear that the model produces clusters primarily based on the stages where they took place. Interestingly however, we also see the influence from the date of the concert, origin of the band, and the genre. Although the presented results are not very strong statistically, we conclude that the model does produce clusters that relate to features of the concerts/bands.

We can describe the produced clusters (1-10) based on their relations to features:
\begin{enumerate}
\item Electronic concerts from the main days of the festival, happening at the three stages (Cosmopol, Gloria, Odeon).
\item Danish bands playing in the warm-up days at Pavilion Junior stage.
\item Various genres from the first days of the main festival from three stages (Cosmopol, Gloria, Odeon).
\item Concerts from the first days of the main festival from Pavilion stage.
\item Mainly concerts from the second (largest) day of the main festival from various stages.
\item Danish and other Nordic bands entirely from the warm-up days.
\item Folk and World bands from the main days of the festival, mostly from the smallest Gloria stage.
\item Bands from the US playing various genres on the last day at different stages.
\item Various bands from the main days playing at Pavilion.
\item Concerts happening on the last day, possibly capturing one-day-ticket participants.
\end{enumerate}

\begin{figure}[!h]
\centering
\includegraphics[width=1\columnwidth]{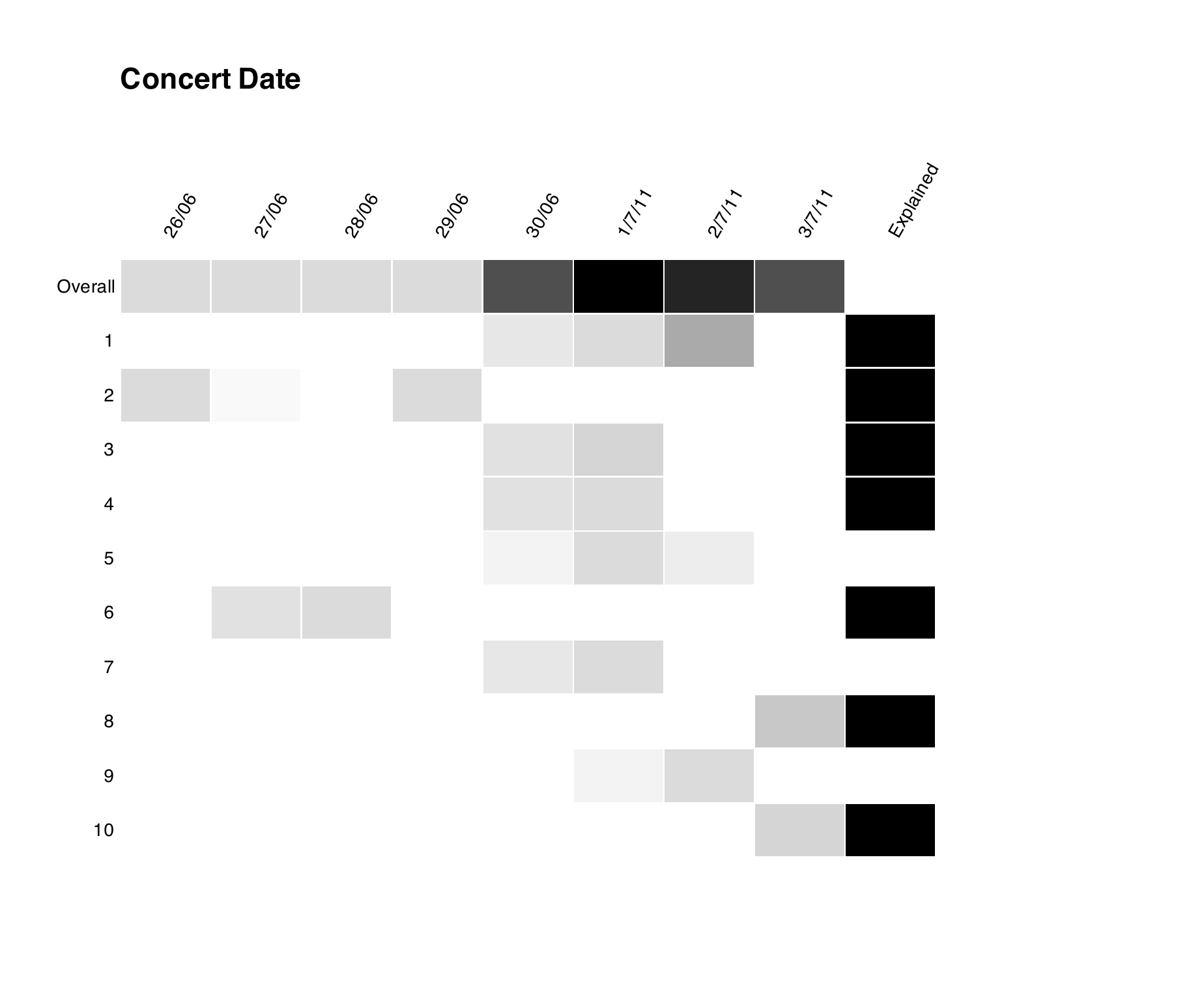}
\caption{Distribution of concert dates in clusters. 7 out of 10 clusters have dates distribution significantly different from the overall.}
\label{fig:clusters_date}
\end{figure}

\begin{figure}[!h]
\centering
\includegraphics[width=1\columnwidth]{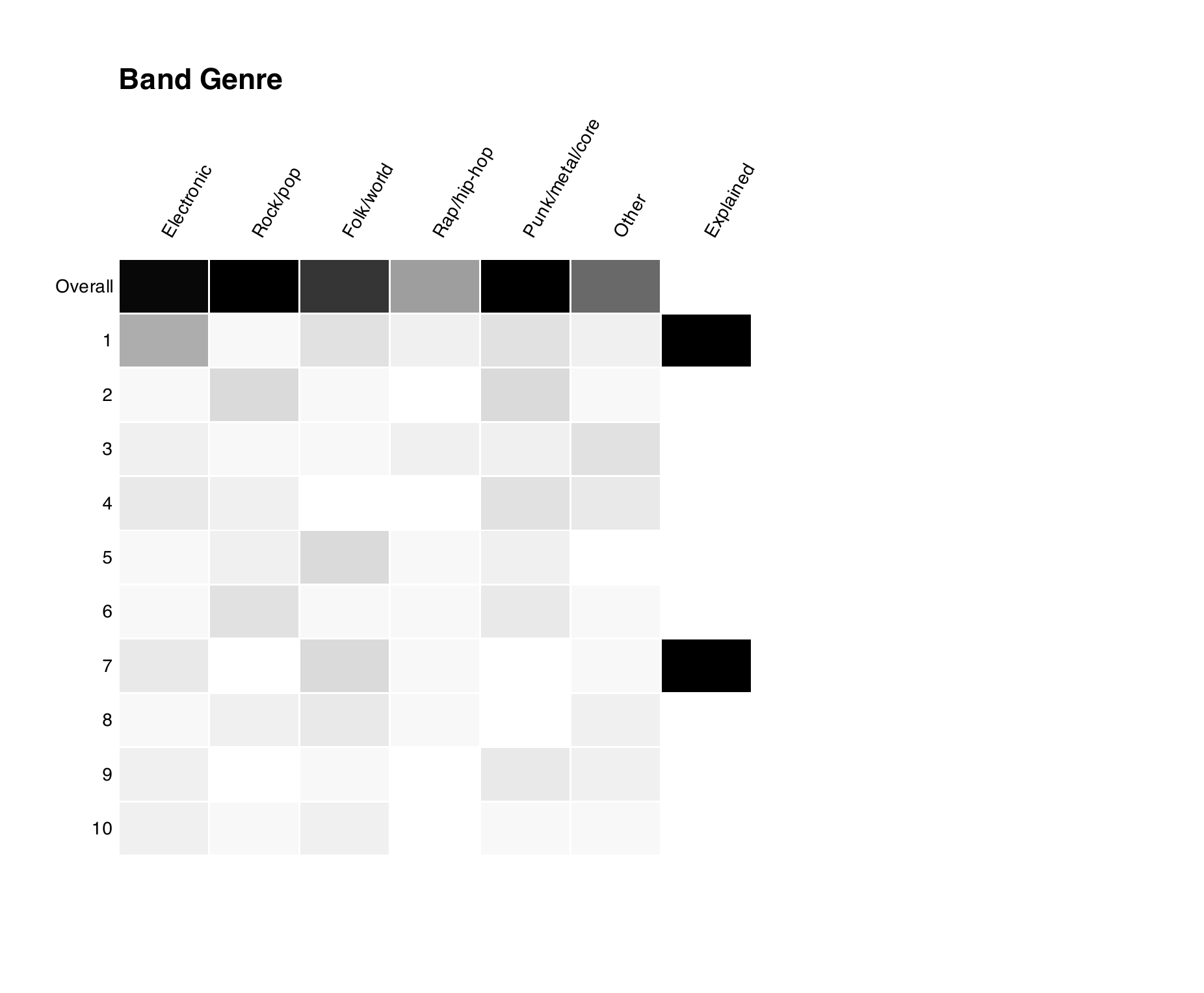}
\caption{Distribution of concert genres in clusters. Two clusters have distribution significantly different from the overall.}
\label{fig:clusters_genre}
\end{figure}

\begin{figure}[!h]
\centering
\includegraphics[width=1\columnwidth]{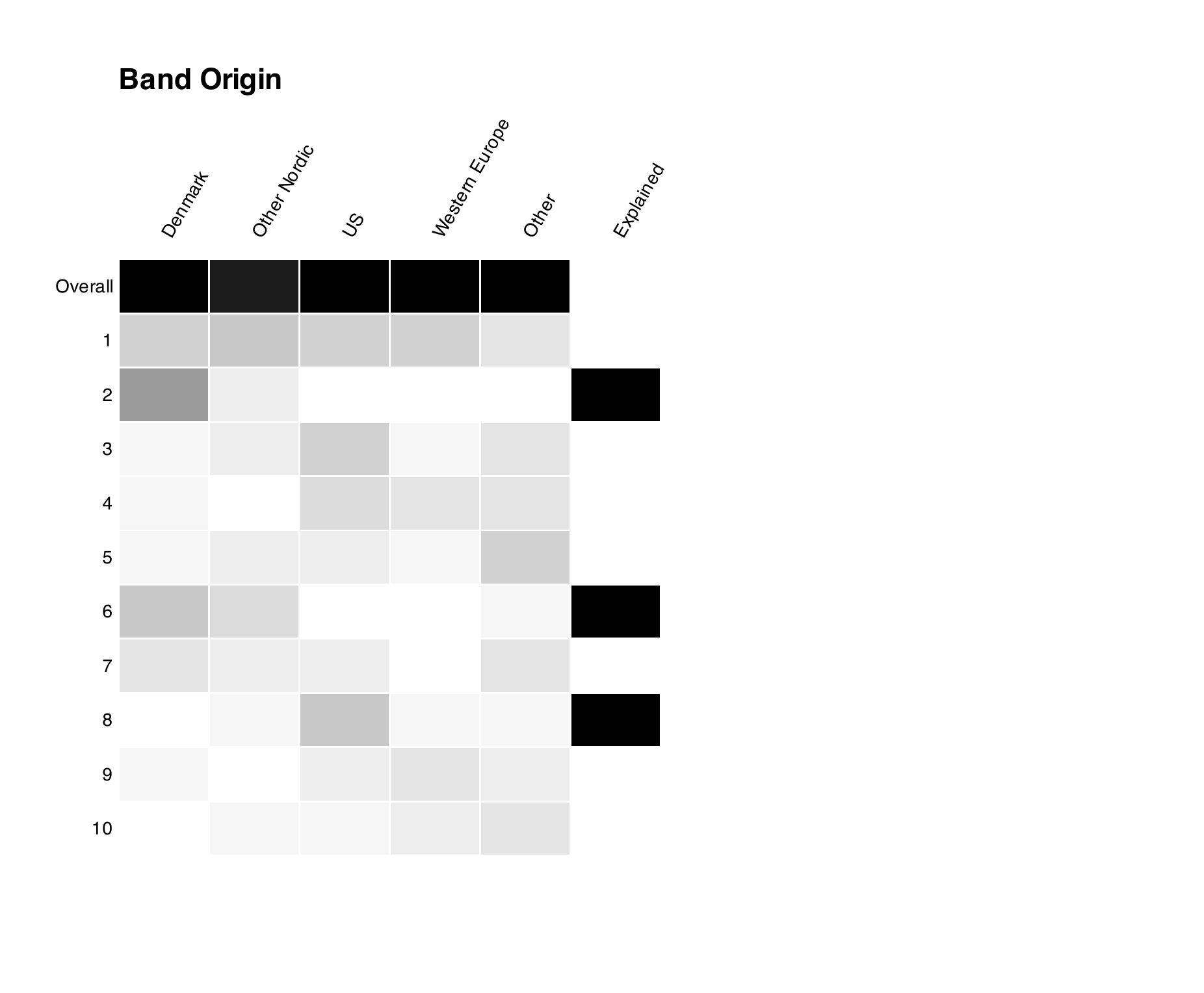}
\caption{Band origin distribution in the clusters. Three clusters show significant grouping of bands: Denmark, Denmark + Other Nordic, and US.}
\label{fig:clusters_origin}
\end{figure}

\begin{figure}[!h]
\centering
\includegraphics[width=1\columnwidth]{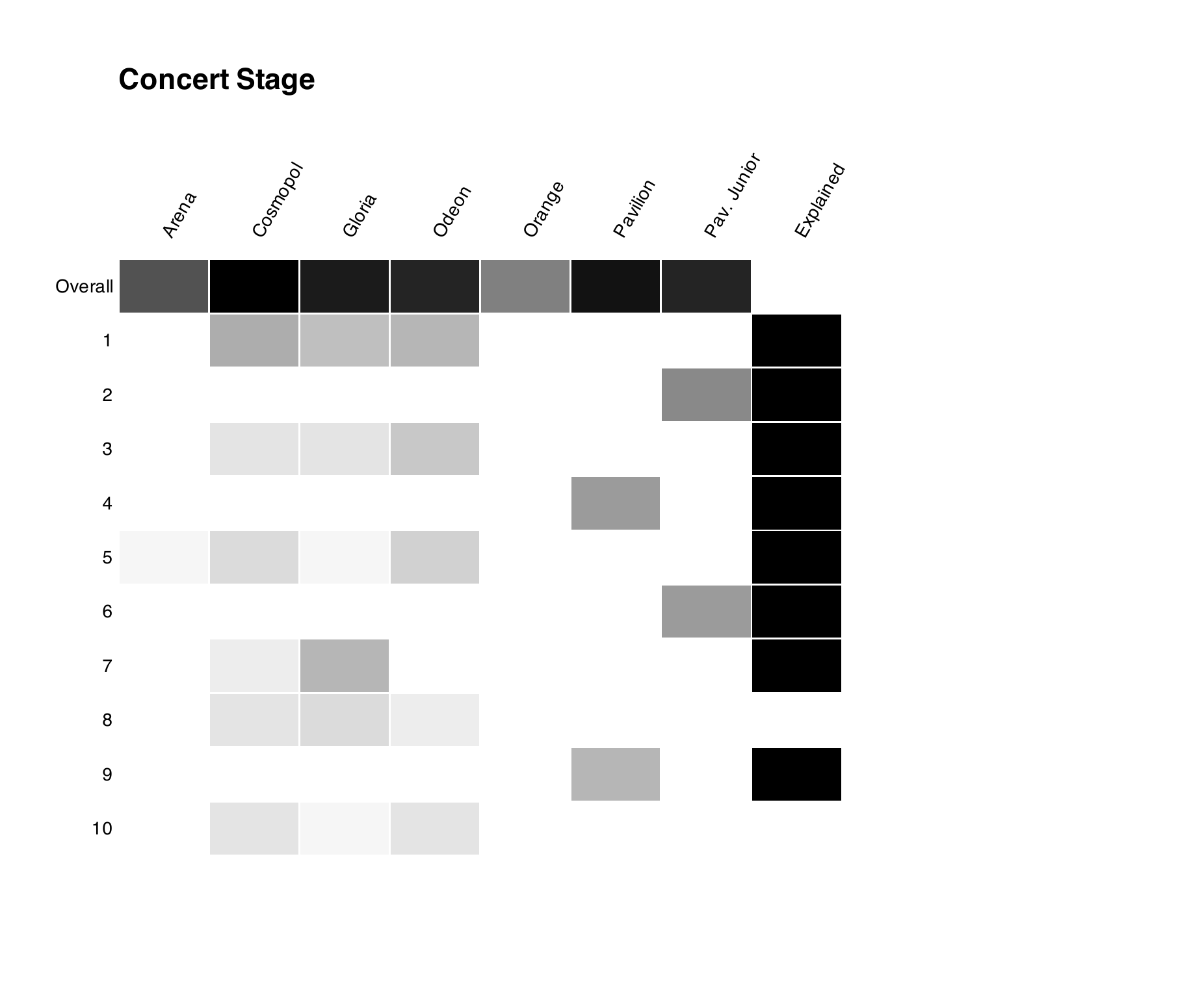}
\caption{Distribution of concert stages. We can notice most of the clusters displaying significant grouping of the concerts according to the stage.}
\label{fig:clusters_stage}
\end{figure}

\begin{figure}[!h]
\centering
\includegraphics[width=1\columnwidth]{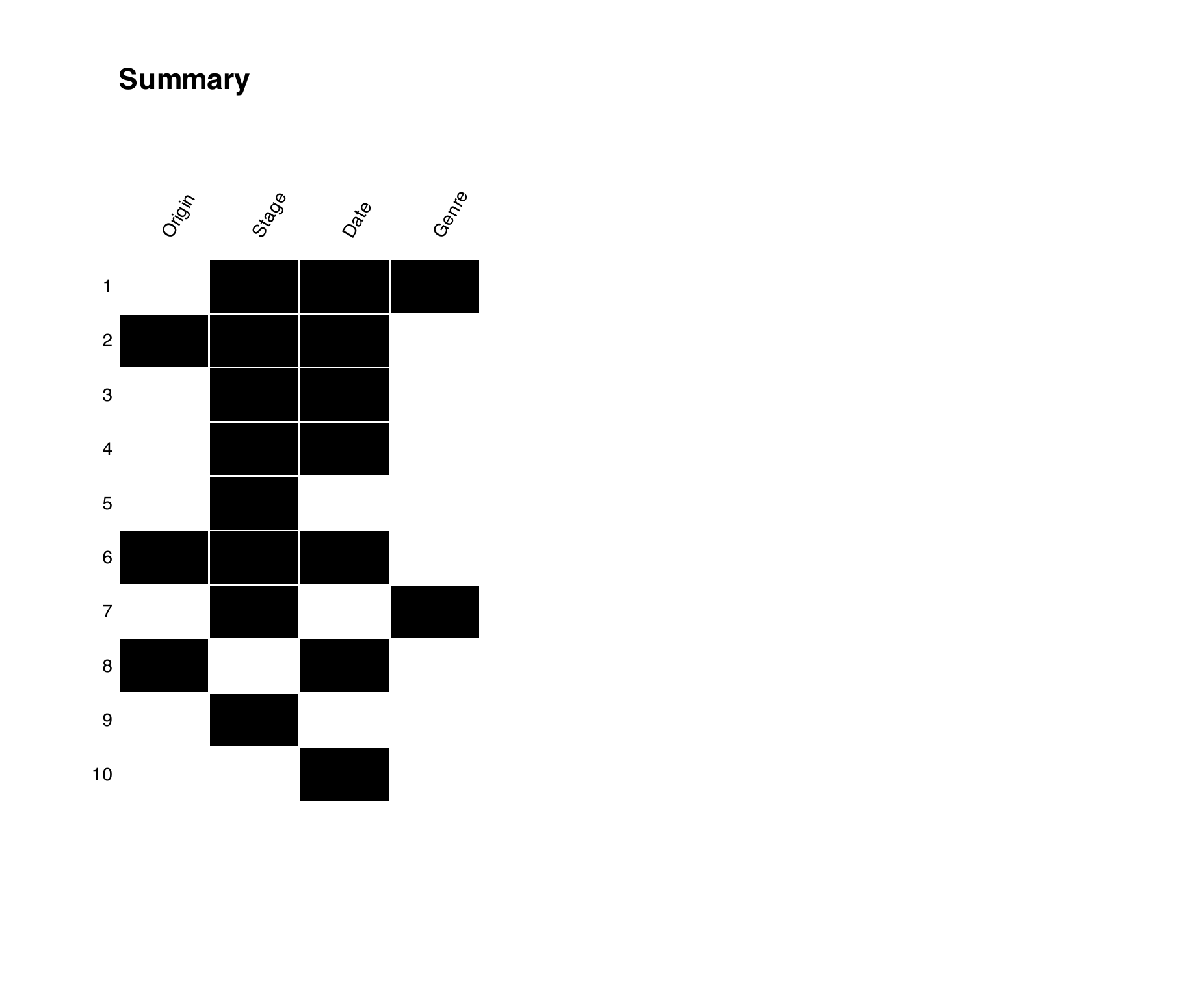}
\caption{Summary of the clusters and features where their distribution is significantly different from overall. }
\label{fig:clusters_summary}
\end{figure}

		\subsubsection{Between cluster link probability matrix}
As shown in Figure~\ref{fig:concert_user}, there are several clusters of participants which show very specific preferences regarding the concerts. For example, participant group 5 (392 persons) only attended concerts from clusters 8, 10, 15, 16, 17, 20, 21. Nearly all of the concerts in these clusters took place on 3rd of July (last day of the festival with major bands performing). Participants from group 4 (475 persons) showed similar preference on that day but they also attended concerts on other days. Participant group 6 (352 persons) behaved like participant group 4 on days other than 3rd of July but showed no interest in the concerts on that day.
Another participant group which shows a clear pattern in concert attendance is group 12 (91 persons) which has high link probabilities with clusters 4, 9, 16, and 24. It occurs that all of the concerts from these clusters took place at the Pavilion stage. 

\section{Discussion}

Our study has demonstrated that discovery of Bluetooth devices at large-scale events can provide interesting insights on participant behavior, group formation, and music preferences. The analysis of the collected Bluetooth data has demonstrated how the spatio-temporal data can reveal underlying structures, when combined with additional contextual metadata describing the concerts and music genres.  In the present study we found that over the duration of the festival 6-7\% of the participants appear to have Bluetooth switched on and in discoverable mode. However, based on the available data it is not possible to conclude on the reasons for this or the actual usage of Bluetooth. Moreover, we were able to observe the distribution of vendors of the discovered devices, but this distribution may not correspond directly to the actual distribution of mobile phones at the festival. In other words, the Bluetooth discoverable devices may not be representative, as for instance most Android-based smartphones only allow time limited discoverability. As such we would expect to observe fewer Android devices in our dataset than there actually are at the Festival. The increasing adaptation of the Android smartphones could perhaps account for the lower penetration of Bluetooth-discoverable devices in the crowd when compared to \cite{versichele2012use}.

The spatio-temporal data allow for analysis of co-occurrences of participants, thereby giving indications of group formation among the festival participants. Furthermore, an advantage of the Bluetooth methodology for doing participant census is that we learn the identity of devices. With this, it is not only possible to estimate the number of people present at different concerts but also determine patterns in the selection of different concert across the entire festival, based on music profiles determined from the spatio-temporal data. Therefore the analysis of this data have provided insights into the underlying structures, that is, the discovery of groups with specific behaviors (music preferences) in terms of choosing concerts. Our analysis shows, that the allocation of artists in terms of stage and day of Festival when they perform is a crucial issue.  We find that many people are not willing to move around the festival area - instead participants tend to spend much of their time around a particular stage. We also show, that for those who do attend concerts at different locations, the country of origin of a band is an important factor when selecting the gigs. Furthermore, we do not find clusters of fans of particular music genre which means the participants are open towards different kinds of performances. Such information can be very valuable for the Festival organizers in the process of booking and allocating bands to stages.

As the collected data was uploaded continuously by the scanners it was possible to create a near real-time visualization of the location of participants at the festival. The real time visualization displayed the activity as the number of unique devices seen in half-hour time windows in different zones of the festival and mapped this information onto a 3D model of the festival area. The rotating 3D model was displayed on a 46 inch monitor located in the so-called {\em Social Zone} of the festival and ran in continuous loops, displaying speed up of activities from the first day until the current moment, see Figure~\ref{fig:viz_setup}. This way of visualizing the activity data allowed for high dynamic of normally slower changing patterns, an easy overview of the festival activity so far, and the possibility of incorporating past data that was only uploaded later (in case scanners did not have a network connection). This setup also allowed us to test the feasibility of obtaining the Bluetooth data in real time using the regular cellular 3G network as a way to build end-user applications on top of the system.

\begin{figure}[!h]
\centering
\includegraphics[width=1.0\columnwidth]{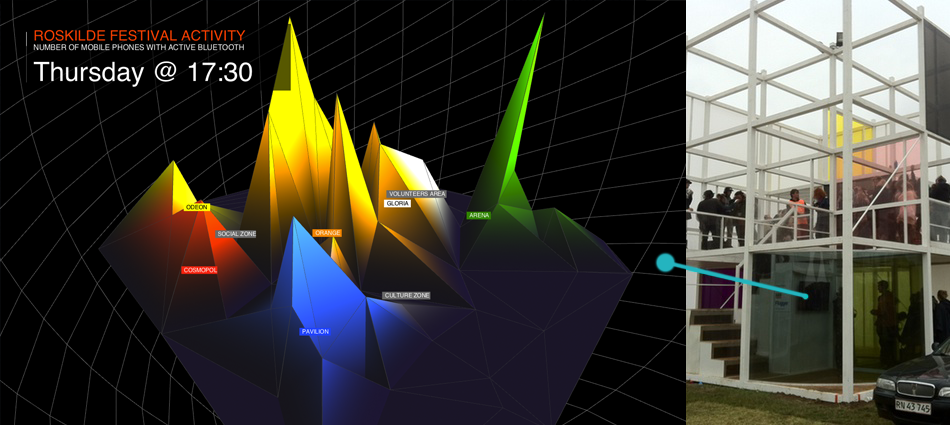}
\caption{The 3D real time animated visualization shown to participants on a large display situated inside a cubic installation that also hosted a silent disco. The 3D model of the festival area was continuously rotating and replaying the visualization of the collected Bluetooth data from the beginning of the festival up to the current moment.}
\label{fig:viz_setup}
\end{figure}

At the festival we were able to observe participants as they experienced the visualization of the Bluetooth data. Initially, they were attracted by the animation, bright colors, and high dynamics, then they subsequently understood what was shown in the visualization. In the setup that was deployed at this festival, the interaction through the 3D visualization of the Bluetooth devices in the festival areas was indirect. The analysis of the data has demonstrated that even more sophisticated participant feedback could be included in such a visualization -- even in real-time. Furthermore, it could allow for more direct interaction through mobile social apps on participant smartphones. For instance to locate groups, participants, or relevant events, as they are happening at the festival. 

As mentioned in the introduction, sensor frameworks for smartphones have received increased attention recently. Future studies could further improve the data collection at a large scale event through the richer datasets that can be obtained from smartphone embedded sensors. By distributing the scanning on multiple client devices the inherent limitation of the present short-range proximity based probing approach may be addressed. In the current setup it is challenging to cover a large physical area in addition to the set of challenges in deploying the system -- including limited availability of power and network in the festival settings. However, a challenge in the distributed scanning approach is the deployment of a sufficient number of client devices in order to obtain sufficient continuous coverage of the area. The initial steps in the direction of distributed Bluetooth scanning were taken by Stopczy\'{n}ski et al.~\cite{stopczynski2013}.

We believe that the results that can be obtained from this Bluetooth probing methodology may also be useful on multiple levels for the festival organizers. The data can help the organizers in assessing participant reactions to the music selection and distribution over the different stages. A more detailed analysis of participant mobility may also help the organizers in planning the layout of the festival area for future festivals.   
\section{Conclusions}

In this paper we have shown that proximity-based Bluetooth sensing is a useful method for obtaining spatio-temporal data in a large-scale event setting. It is possible to analyze the data, accounting for sparsity and missing data using mathematical models and discover meaningful patterns of participant behavior, including mobility, group formation, and music preferences. We have also demonstrated the feasibility of capturing Bluetooth data from a large crowd and visualize the resulting spatio-temporal data in real time. Finally, we have proposed how the Bluetooth probing methodology may serve as a framework for creating future mobile social interaction applications for such large-scale events. 
\vfill
\section{Acknowledgment}

We would like to thank the Roskilde Festival organizers. Also thanks to Nokia for partly sponsoring the mobile phones used as part of the study. Finally thanks to Krzysztof Siejkowski, Marcin Ignac, and S$\o$ren Rosenbak.


%
%
%
%
\balance

\small
\bibliographystyle{acm-sigchi}
\bibliography{sample}

\begin{thebibliography}{10}

\bibitem{aharony2011social}
Aharony, N., Pan, W., Ip, C., Khayal, I., and Pentland, A.
\newblock Social fmri: Investigating and shaping social mechanisms in the real
  world.
\newblock {\em Pervasive and Mobile Computing\/} (2011).

\bibitem{bensky2007wireless}
Bensky, A.
\newblock {\em Wireless positioning technologies and applications}.
\newblock Artech House, Inc., 2007.

\bibitem{bullock2010automated}
Bullock, D., Haseman, R., Wasson, J., and Spitler, R.
\newblock Automated measurement of wait times at airport security.
\newblock {\em Journal of the Transportation Research Board 2177}, -1 (2010),
  60--68.

\bibitem{Crandall28122010}
Crandall, D.~J., Backstrom, L., Cosley, D., Suri, S., Huttenlocher, D., and
  Kleinberg, J.
\newblock Inferring social ties from geographic coincidences.
\newblock {\em Proc. of the National Academy of Sciences 107}, 52 (2010),
  22436--22441.

\bibitem{Dahl2005}
Dahl, D.~B.
\newblock Sequentially-allocated merge-split sampler for conjugate and
  nonconjugate {D}irichlet process mixture models.
\newblock Tech. rep., Texas A\&M University, 2005.

\bibitem{eagle2006reality}
Eagle, N., and Pentland, A.
\newblock Reality mining: sensing complex social systems.
\newblock {\em Personal and Ubiquitous Computing 10}, 4 (2006), 255--268.

\bibitem{gonzalez2008understanding}
Gonzalez, M., Hidalgo, C., and Barabasi, A.
\newblock Understanding individual human mobility patterns.
\newblock {\em Nature 453}, 7196 (2008), 779--782.

\bibitem{hansen2009location}
Hansen, J., Alapetite, A., Andersen, H., Malmborg, L., and Thommesen, J.
\newblock Location-based services and privacy in airports.
\newblock {\em Human-Computer Interaction--INTERACT 2009\/} (2009), 168--181.

\bibitem{hansen2011}
Hansen, T., Morup, M., and Hansen, L.
\newblock Non-parametric co-clustering of large scale sparse bipartite networks
  on the gpu.
\newblock In {\em IEEE Int. Workshop on Machine Learning for Signal Processing
  (MLSP)} (2011), 1 --6.

\bibitem{haseman2010real}
Haseman, R., Wasson, J., and Bullock, D.
\newblock Real time measurement of work zone travel time delay and evaluation
  metrics using bluetooth probe tracking.
\newblock {\em Journal of the Transportation Research Board\/} (2010).

\bibitem{hay2009bluetooth}
Hay, S., and Harle, R.
\newblock Bluetooth tracking without discoverability.
\newblock {\em Location and Context Awareness\/} (2009), 120--137.

\bibitem{hui2005pocket}
Hui, P., Chaintreau, A., Scott, J., Gass, R., Crowcroft, J., and Diot, C.
\newblock Pocket switched networks and human mobility in conference
  environments.
\newblock In {\em Proceedings of the 2005 ACM SIGCOMM workshop on
  Delay-tolerant networking}, ACM (2005), 244--251.

\bibitem{ieeeoui}
IEEE.
\newblock {P}ublic {OUI} and {C}ompany {ID} {A}ssignments.
\newblock \url{http://standards.ieee.org/develop/regauth/oui/}.

\bibitem{Jain2004}
Jain, S., and Neal, R.~M.
\newblock A split-merge markov chain monte carlo procedure for the dirichlet
  process mixture model.
\newblock {\em Journal of Computational and Graphical Statistics 13}, 1 (2004),
  158--182.

\bibitem{jensen2010estimating}
Jensen, B., Larsen, J., Jensen, K., Larsen, J., and Hansen, L.
\newblock Estimating human predictability from mobile sensor data.
\newblock In {\em IEEE Int. Workshop on Machine Learning for Signal Processing
  (MLSP)} (2010), 196--201.

\bibitem{kelly2010minimal}
Kelly, D.
\newblock {\em Minimal Infrastructure Radio Frequency Home Localisation
  Systems}.
\newblock PhD thesis, National University of Ireland, 2010.

\bibitem{Kemp2006}
Kemp, C., Tenenbaum, J.~B., Griffiths, T.~L., Yamada, T., and Ueda, N.
\newblock Learning systems of concepts with an infinite relational model.
\newblock In {\em Proc. of the National AAAI Conf. on Artificial Intelligence}
  (2006).

\bibitem{kiukkonen2010towards}
Kiukkonen, N., Blom, J., Dousse, O., Gatica-Perez, D., and Laurila, J.
\newblock Towards rich mobile phone datasets: Lausanne data collection
  campaign.
\newblock {\em Proc. ICPS\/} (2010).

\bibitem{Kostakos2010}
Kostakos, V., O'Neill, E., Penn, A., Roussos, G., and Papadongonas, D.
\newblock Brief encounters: Sensing, modeling and visualizing urban mobility
  and copresence networks.
\newblock {\em ACM Trans. Comput.-Hum. Interact. 17}, 1 (Apr. 2010), 2:1--2:38.

\bibitem{maslov2004detection}
Maslov, S., Sneppen, K., and Zaliznyak, A.
\newblock Detection of topological patterns in complex networks: correlation
  profile of the internet.
\newblock {\em Physica A: Statistical Mechanics and its Applications 333\/}
  (2004), 529--540.

\bibitem{montoliu2010discovering}
Montoliu, R., and Gatica-Perez, D.
\newblock Discovering human places of interest from multimodal mobile phone
  data.
\newblock In {\em Proceedings of the 9th International Conference on Mobile and
  Ubiquitous Multimedia}, ACM (2010), 12.

\bibitem{ONeill2006}
O'Neill, E., Kostakos, V., Kindberg, T., Schiek, A. F.~g., Penn, A., Fraser,
  D.~S., and Jones, T.
\newblock Instrumenting the city: developing methods for observing and
  understanding the digital cityscape.
\newblock In {\em Proceedings of the 8th international conference on Ubiquitous
  Computing}, UbiComp'06, Springer-Verlag (Berlin, Heidelberg, 2006), 315--332.

\bibitem{peterson2006bluetooth}
Peterson, B.~S., Baldwin, R.~O., and Kharoufeh, J.~P.
\newblock Bluetooth inquiry time characterization and selection.
\newblock {\em IEEE Transactions on Mobile Computing 5}, 9 (2006), 1173--1187.

\bibitem{pitman2006combinatorial}
Pitman, J.
\newblock {\em Combinatorial stochastic processes}, vol.~1875.
\newblock Springer-Verlag, 2006.

\bibitem{song2010limits}
Song, C., Qu, Z., Blumm, N., and Barab{\'a}si, A.
\newblock Limits of predictability in human mobility.
\newblock {\em Science 327}, 5968 (2010), 1018.

\bibitem{stange2011analytical}
Stange, H., Liebig, T., Hecker, D., Andrienko, G., and Andrienko, N.
\newblock Analytical workflow of monitoring human mobility in big event
  settings using bluetooth.
\newblock In {\em Proc. of the 3rd ACM SIGSPATIAL Int.l Workshop on Indoor
  Spatial Awareness}, ACM (2011), 51--58.

\bibitem{stopczynski2013}
Stopczynski, A., Larsen, J., Lehmann, S., L., D., and M., F.
\newblock Participatory {B}luetooth {S}ensing: {A} {M}ethod for {A}cquiring
  {S}patio-{T}emporal {D}ata about {P}articipant {M}obility and {I}nteractions
  at {L}arge {S}cale {E}vents.
\newblock In {\em Pervasive Computing and Communications Workshops, 2013.
  PerCom Workshops '13} (2013).

\bibitem{steinbach2006}
Tan, P.-N., Steinbach, M., and Kumar, V.
\newblock {\em Introduction to Data Mining, (First Edition)}.
\newblock Addison-Wesley Longman Publishing Co., Inc., Boston, MA, USA, 2006.

\bibitem{versichele2010potential}
Versichele, M., Delafontaine, M., Neutens, T., and Van~de Weghe, N.
\newblock Potential and implications of bluetooth proximity-based tracking in
  moving object research.
\newblock In {\em 1st Int. workshop on movement pattern analysis (MPA) in conj.
  with the 6th Int. conf. on Geographic Information Science} (2010).

\bibitem{versichele2012use}
Versichele, M., Neutens, T., Delafontaine, M., and Van~de Weghe, N.
\newblock The use of bluetooth for analysing spatiotemporal dynamics of human
  movement at mass events: A case study of the ghent festivities.
\newblock {\em Applied Geography 32}, 2 (2012), 208--220.

\bibitem{Xu2006}
Xu, Z., Tresp, V., Yu, K., and Kriegel, H.-P.
\newblock Learning infinite hidden relational models.
\newblock {\em Uncertainty in Artificial Intelligence (UAI2006)\/} (2006).

\end{thebibliography}
\end{document}